\input{aipcheck}

\documentclass[
    ,final            
  ]
  {aipproc}

\layoutstyle{6x9}

\begin{document}

\title[H$\alpha$ Surveys] {Discovering Interacting Binaries with H$\alpha$ Surveys}
\author{Andrew Witham} {address={School of Physics \& Astronomy, University of Southampton, Highfield, Southampton, SO17 1BJ, UK}}
\author{Christian Knigge} {address={School of Physics \& Astronomy, University of Southampton, Highfield, Southampton, SO17 1BJ, UK}}
\author{Janet Drew} {address={Astrophysics Group, Department of Physics, Imperial College London, Exhibition Road, London SW7 2AZ, UK}}
\author{Paul Groot} {address={Department of Astrophysics, University of Nijmegen, P.O. Box 9010, NL - 6500 GL Nijmegen, NL}}
\author{Robert Greimel} {address={Isaac Newton Group, Apartado de Correos, 321, 38700 Santa Cruz de La Palma, Canary Islands, ESP}}
\author{Quentin Parker} {address={Department of Physics, Macquarie University, NSW 2109, Australia}, altaddress={Anglo-Australian Observatory, PO Box 296, Epping NSW 1710, Australia}}

\begin{abstract}
A deep (R $\sim$ 19.5) photographic H$\alpha$ Survey of the southern Galactic
Plane was recently completed using the UK Schmidt Telescope at the AAO.
In addition, we have recently started a similar, CCD-based survey of the
northern Galactic Plane using the Wide Field Camera on the INT. Both surveys aim to provide
information on many types of emission line objects, such as planetary
nebulae, luminous blue variables and interacting binaries.

Here, we focus specifically on the ability of H$\alpha$ emission line surveys
to discover cataclysmic variables (CVs).  Follow-up observations have already begun,
and we present initial spectra of a candidate CV discovered by these surveys.  We also present results from analyzing the properties of known CVs in the Southern Survey. By calculating the recovery rate of these objects, we estimate the efficiency of H$\alpha$-based searches in
finding CVs.

\end{abstract}

\maketitle


\section{Introduction}
Binary evolution theory predicts a
large (but currently undetected) population of short-period, faint cataclysmic cariables (CVs) (see for example Kolb 1993 \cite{Kolb93}; and Howell, Rappaport, \& Politano 1997 \cite{Howe97}).
Such systems should exhibit particularly strong Balmer emission (Patterson 1984 \cite{Patt84}), making
H$\alpha$ surveys an ideal way to find them, if they exist.  Previous H$\alpha$ surveys (see for example Kohoutek and Wehmeyer 1997 \cite{kandw97}; and Gaustad \textit{et al} 2001 \cite{gaus01}) have become incomplete
at bright magnitudes or have been of low spatial resolution.  Below, we briefly describe two new photometric H$\alpha$ surveys which are able to push to deeper magnitudes with good resolution and will allow for a significant population of new H$\alpha$ emitters to be discovered, including many Cataclysmic Variables.

\section{H$\alpha$ Surveys}
The galactic plane is being observed in $H\alpha$ by two Surveys: (i)
The INT/WFC Photometric H$\alpha$ Survey of the Northern Galactic Plane (IPHAS), and (ii) the AAO/UKST Southern H$\alpha$ Survey (SHS).
IPHAS is a CCD survey which combines photometry in the bands H$\alpha$, r and i from images obtained with the Wide Field Camera (WFC) on
the Isaac Newton Telescope (INT).  The WFC provides a field of view of 34'x34' and the resolution of the accepted photometry is $\leq$ 1.7".
At the time of writing, IPHAS is $\sim$ 1/3 complete and when finished will cover the whole of the Northern galactic plane with latitudes |b| $<\sim5^{\circ}$ down to $\sim$ 19.5 mags in r and i.  Further information can be found at \url{http://astro.ic.ac.uk/Research/Halpha/North} and in a forthcoming paper by Drew \textit{et al} (in preparation).

The SHS is a photographic survey which, using the UK Schmidt Telescope (UKST), has provided images of 233 galactic fields at four degree centres.
Fields were observed in two bands (H$\alpha$ and short red $\simeq$ R) with a spatial resolution of $\sim$ 1".
This has resulted in a survey that is complete to $\sim$ 19.5 mags in R and has an area of coverage defined by $-75 < Dec < +2.5$ and |b| $<\sim10^{\circ}$.
The data can be accessed via the SuperCOSMOS website at \url{http://www-wfau.roe.ac.uk/sss/halpha} and additional information can be found in Parker \& Phillipps (1998) \cite{pandp98}

\section{Spectroscopic Follow-Up Of IPHAS}
\begin{figure}
{\includegraphics[width=9cm,angle=270,clip=true]{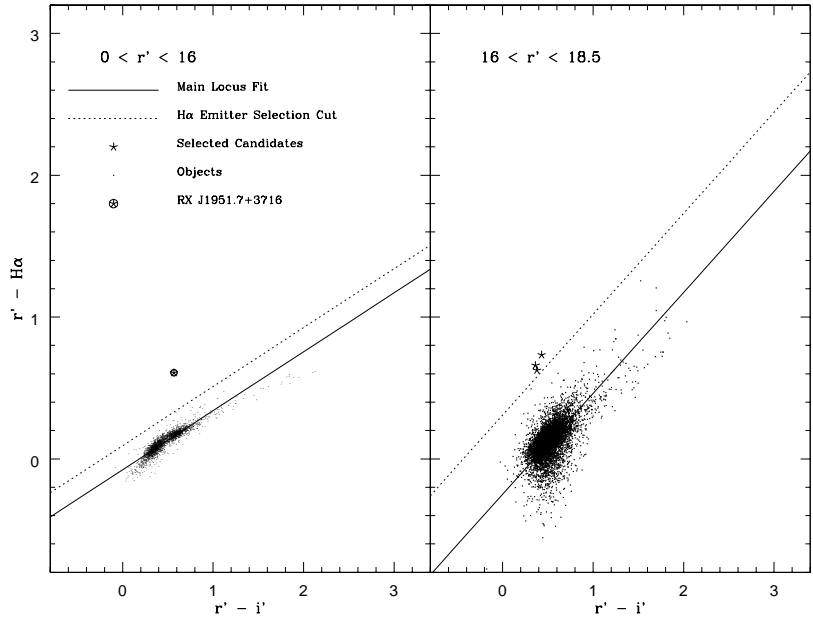}}
\caption{\label{fig:d} Illustration of our selection strategy.  The plots compromise two r magnitude bins containing the objects from one IPHAS field.  A H$\alpha$ excess of 0.1mag corresponds to an equivalent width of $\sim$10\AA. Solid lines are linear fits to the stellar loci, the dotted lines indicate the selection cut.  The known CV RX J1951.7+3716 is present within this field and would be selected as it lies above the cut.}

\end{figure}
\begin{figure}
{\includegraphics[width=9cm,angle=270,clip=true]{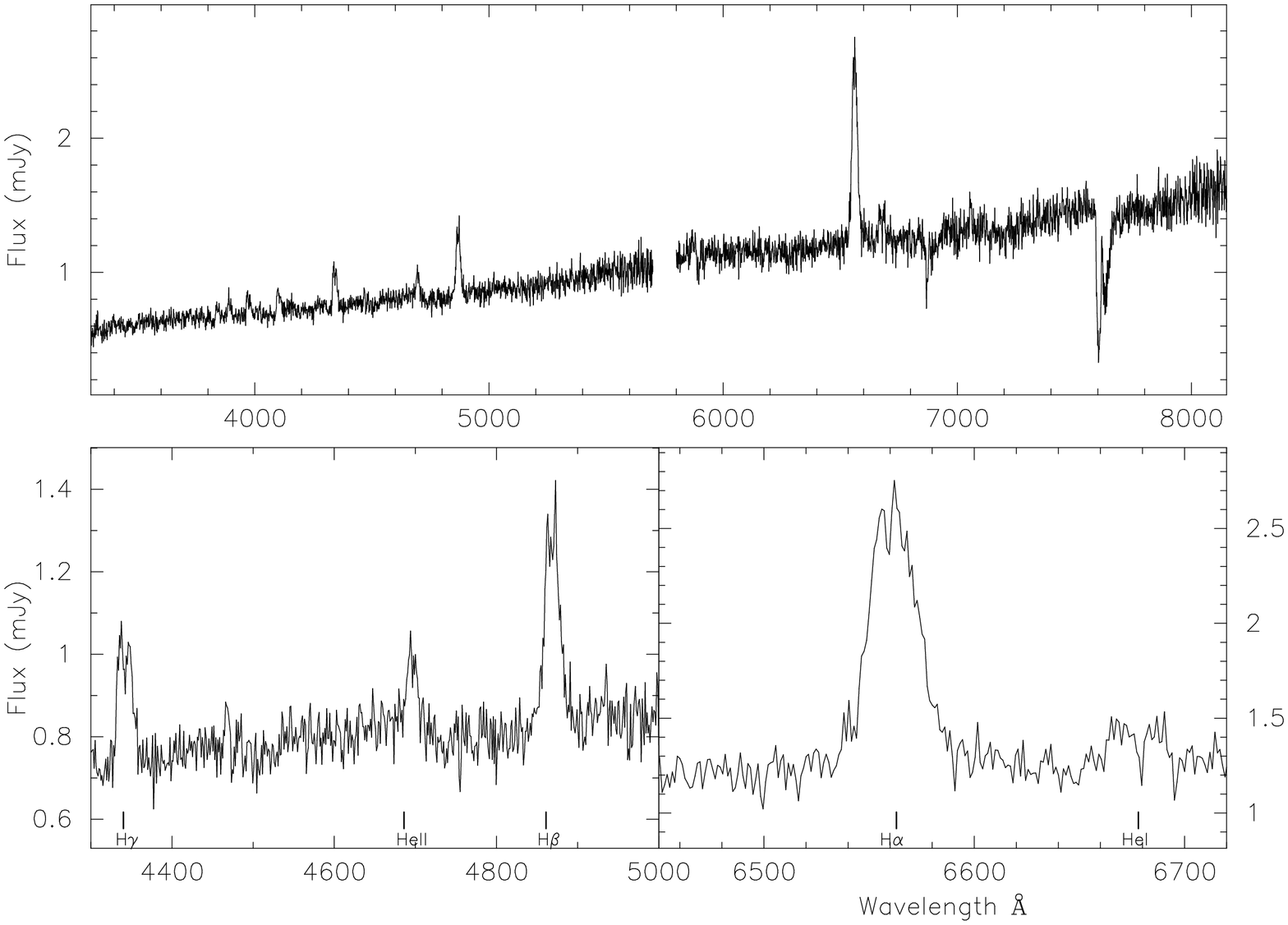}}
\caption{\label{fig:e} WHT/ISIS spectra of a newly discovered CV candidate.  The dispersion is
0.86\AA/pixel in the blue and 1.49\AA/pixel in the red.  Based on the broad and strong H$\alpha$ emission line
and the presence of HeII emission, this is a CV candidate.}
\end{figure}
IPHAS has the advantage of having far fewer spurious $H\alpha$ emitters than the SHS because of the photographic nature of the SHS.  Defining selection criteria for follow-up is therefore less problematic with IPHAS.  Initial selection of potential new CVs for follow-up was done on the basis that we expect many to be H$\alpha$ emitters and to show a large H$\alpha$ excess.  These CVs can thus be expected to lie above the main stellar locus in colour-colour plots of r-H$\alpha$ vs r-i.

Spectroscopic follow-up of IPHAS is in progress and has already lead to the discovery of three new CV candidates.  Figure \ref{fig:d} illustrates the selection criteria used to pick objects for follow-up, while Figure \ref{fig:e} shows the resulting spectrum of one of the
new CV candidates.  Classification of the new CV candidates will be done using time-resolved
spectroscopy and photometry.



\section{Recovery Rate of Known CVs In The South}
\begin{figure}
{\includegraphics[width=8.3cm,angle=0,clip=true]{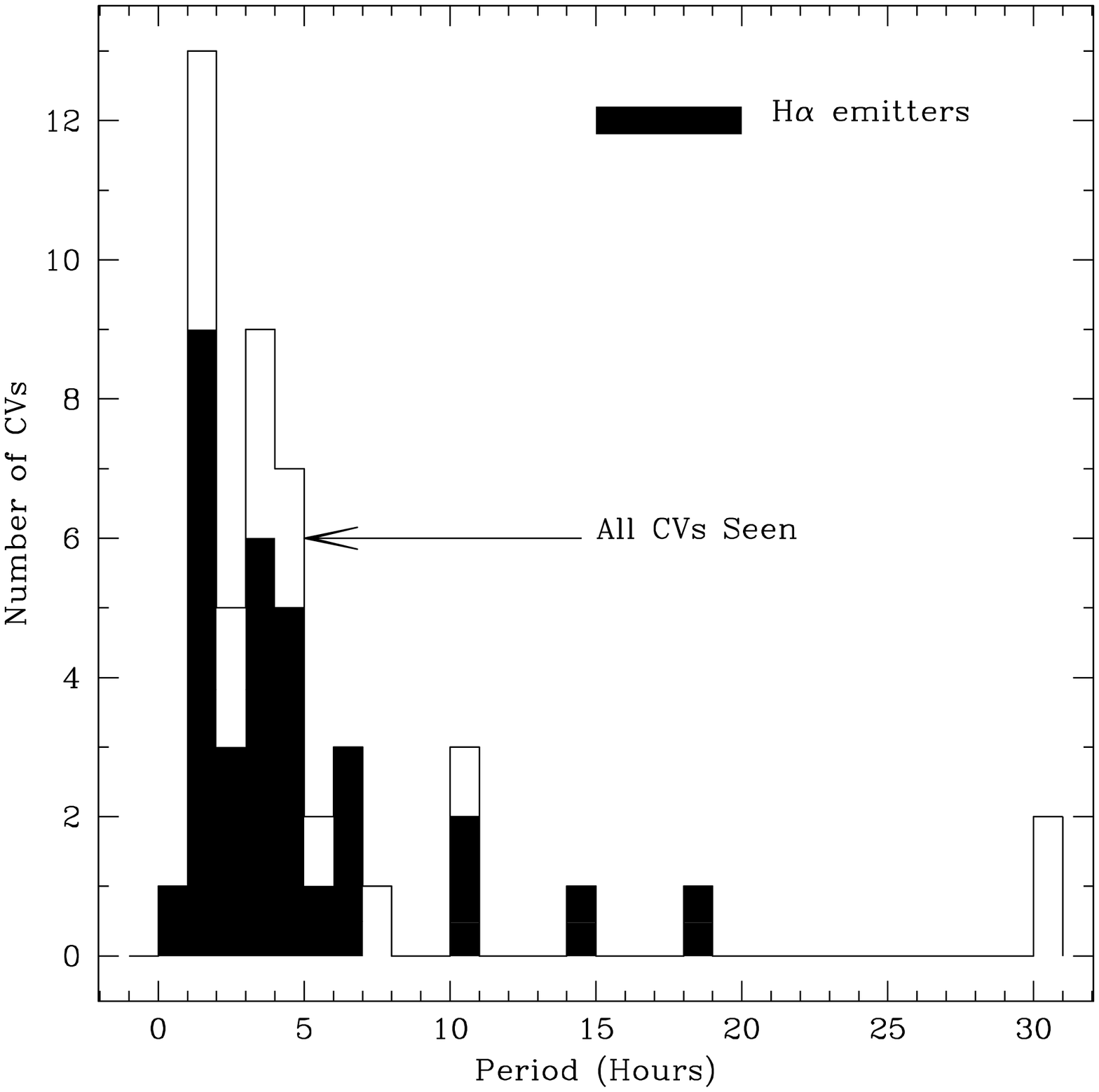}}
\caption{\label{fig:b}The period distribution of known CVs classified as emitters (black) overlaid on the distribution of the known CVs with good photometry in the survey (white).}
\end{figure}


We have analyzed the properties of the CVs contained in the Ritter \& Kolb (2003) \cite{randk03} catalog that
fall within the SHS area.  Using R-H$\alpha$ vs R band colour-magnitude plots we classify as "emitters" those CVs that would have been selected
for spectroscopic follow-up with 6df \textit{i.e.} 5 objects per square degree.
The result is that 52\% of the sample would be recovered by the Southern Survey.  Furthermore, if consideration
is limited to objects with good photometry in the survey, the recovery rate is even higher at 62\%.  The histogram in Figure \ref{fig:b} shows how the recovery rate depends on the CV orbital period.  Somewhat surprisingly, there is no obvious bias, in favour of detecting faint, short period CVs.

The fact that many of the CVs (including several novae) appear blue as well as having a H$\alpha$ excess can be used to define more stringent selection cuts; however this could have the negative effect of introducing a further selection bias into the CV populations discovered with H$\alpha$ surveys.
A more thorough discussion of the
recovery rate of the known population of CVs in both IPHAS and the SHS will presented by Witham \textit{et al} (in preparation).
Overall the recovery rate is very encouraging and the surveys have a great chance of finding many new CVs.



\end{document}